\documentclass[authoryear,preprint,11pt]{elsarticle} 
\journal{-} 

\usepackage{amsmath}
\usepackage{color}
\usepackage{natbib}
\usepackage{graphicx}
\usepackage{subfigure}
\usepackage{multirow}

\def\gm{{\rm\,g}}
\def\s{{\rm\,s}}
\def\m{{\rm\,m}}
\def\cm{{\rm\,cm}}
\def\km{{\rm\,km}}
\def\yr{\rm\,yr}
\def\mum{\,\mu{\rm m}}
\def\Rs{\,R_s}

\def\Op{\rm\,O^+}
\def\Opp{\rm\,O_2^+}

\DeclareMathOperator\erf{erf} 
\providecommand{\abs}[1]{\lvert#1\rvert} 

\begin{document}

\begin{frontmatter}
    \title{Dust in the arcs of Methone and Anthe}
    \author{Kai-Lung~Sun, Martin~Sei\ss, Frank~Spahn}
    \address{Karl-Liebknecht-Str. 24/25, 14476 Potsdam Golm}

    \begin{abstract}
        Methone and Anthe are two tiny moons (with diameter $<3\km$) in the inner part of Saturn's E ring. Both moons are embedded in an arc of dust particles. To understand the amount of micron-sized dust and their spatial distribution in these arcs, we model the source, dynamical evolution, and sinks of these dust in the arc.
        We assume hypervelocity impacts of micrometeoroids on the moons as source of these dust \citep{Hedman:2009kt}, the so called impact-ejecta process \citep{Krivov:2003ku,Spahn:2006dh}.
        After ejecting and escaping from the moons, these micron-sized particles are subject to several perturbing forces, including gravitational perturbation from Mimas, oblateness of Saturn, Lorentz force, solar radiation pressure, and plasma drag.
        Particles can be either confined in the arcs due to corotational resonance with Mimas, as their source moons \citep{Spitale:2006dk, Cooper:2008hl, Hedman:2009kt}, or pushed outward by plasma drag. Particle sinks are recollisions with the source moon, collision with other moons, or migrate out of the radial zone of interest. In addition to that, the upper limit of the particle lifetimes are controlled by plasma sputtering, which decreases the particle size in a rate of order 1$\mum$ radius every 100 years \citep{Johnson:2008jz}.
        Our simulation results show that ejecta from both moons can form the arcs of maximal optical depths $\tau$ in the order of $10^{-8} - 10^{-6}$, although the absolute amount of dust have uncertainties.
        We also find out the longitudinal extension of the arcs in our simulation are consistent with observation and the theory.
        Smaller particles are more likely to escape the arc because of the stronger influence of plasma drag. On the other hand, large particles can stay in arcs for longer time and therefore are more likely to collide with the source moon. For both arcs about 18-20\% collide with Enceladus, ${\sim}80$\% drift outward or have very high eccentricity, and less than 0.1\% collide with source moons, most of which are large particles.
        Simulation results also show the heliotropic behavior of dust introduced by solar radiation pressure \citep{Hedman:2010ie}.
    \end{abstract}
\end{frontmatter}  
%
\section{Introduction}
Methone, Anthe, Pallene are three tiny moons located at the inner edge of Saturn's E ring, located between the massive moons Mimas (${\sim}3.08\Rs$, $\Rs$=60,268km is Saturn radius) and Enceladus (${\sim}3.95\Rs$). Observations by the Cassini Imaging Science Subsystem (ISS) showed that dusty ring-arcs envelop the moons Methone and Anthe, and a torus exists in the vicinity of Pallene's orbit \citep{Hedman:2009kt}.

Although the mean radii of these moons are less than 3$\km$ \citep{Thomas:2013bz}, they are the most likely sources of these arcs through the impact-ejecta process, which produces dust ejecta caused by impacts of interplanetary micrometeoroids on these moons \citep{Krivov:2003ku,Spahn:2006dh,Hedman:2009kt}. Due to the small size of the moons, the escape speed is small so that nearly all the ejecta can escape from the moon.

Methone occupies both the radial regions of Mimas 15:14 corotation eccentricity resonance (CER) and Mimas 15:14 outer Lindblad resonance, while Anthe is librating inside Mimas 11:10 CER \citep{Spitale:2006dk,Cooper:2008hl,Hedman:2009kt}.
These resonances can trap dust particles released from the surface of these moons, leading to dust arcs in their vicinity. Theoretically, the maximal librating longitude is $360^\circ/15=24^\circ$ for the Methone arc, and $360^\circ/11\approx32^\circ$ for the Anthe arc. The observed length of the Methone arc is about 10$^\circ$, while it is about 20$^\circ$ for Anthe, which means that both arcs must be confined by the nearby resonances \citep{Hedman:2009kt}.
An analog to the Anthe and Methone arcs can be found in the Saturn's G ring arc, which also confined by a resonance with Mimas and also contain a tiny moon \citep{Hedman:2009kt,Hedman:2010iq}.

Although lack of observation, if these arcs are similar to other faint rings like the Saturn G ring and is maintained by impact-ejecta process, then the arcs should contain significant amount of micron-sized particles.
The small grain size implies that several perturbing forces are dominating the shape of these arcs simultaneously. For example, the interaction of solar radiation pressure, oblateness of Saturn, and Lorentz force can result in a heliotropic shape of the ring (an elliptic ring with pericenter or apocenter in sun direction) as observed in the E ring \citep{, Hedman:2012do} and the Charming ringlet in the Cassini Division \citep{Hedman:2010ie}.
Nevertheless, due to the arc structure we see always just a part of the heliotropic ring.
Further, plasma drag pushes E ring micron-sized particles outwards at a rate of order 1$\Rs$ per Saturn year \citep{Dikarev:1999uv, Horanyi:2008ck}.

This paper is organized as follows:
We introduce the impact-eject process, the dynamics, and the sinks of particles in section 2 for modeling, and the simulation results are presented in section 3. Especially we will discuss the heliotropic behavior and the interaction between drag force and resonances in the arc. Discussions and conclusion are given in section 4.

\section{Methods}
\label{sec:methods}
To understand the spatial distribution and amount of dust near Methone and Anthe, we have to model the sources, dynamics, and sinks. As proposed by \citet{Hedman:2009kt}, we assume dust to be produced through the impact-ejecta process \citep{Krivov:2003ku, Spahn:2006dh}, which is generated by the hypervelocity impacts of interplanetary dust on the surfaces of Methone and Anthe.
The dynamics includes several perturbing forces that are summarized in Eq.~\eqref{eq:motion}.
Finally, dust sinks include collisions with moons, paricle erosion due to sputtering, and orbital evolution via plasma drag.

\subsection{Impact-ejecta process}
\label{sec:impeje}
A single hypervelocity impact of a interplanetary dust particle (IDP) on the surface of an atmosphereless body can create a huge amount of ejecta. The mass production rate is given by
\begin{equation}
    M^+ = F_{imp} Y S
    \label{eq:mplus}
\end{equation}
the product of the IDP mass flux $F_{imp}$, the yield $Y$ is the ratio of ejected mass to the projectile mass, and the cross section of target moon $S$.
Both Methone and Anthe are so small that the initial ejecta speeds usually exceed the escape speeds of the moons (as shown in Table.~\ref{tab:n_plus}), so almost all ejecta escape Methone and Anthe.
The IDP mass flux $F_{imp}$ at the Saturn environment is not yet well determined by observations, we adapt the value from \citet{Krivov:2003ku}, who used the \citet{Divine:1993ip} model. This value may be overestimated by an order of magnitude and thus needs a further update. The yield $Y$ is also quite uncertain, it's value depends sensitively on surface properties, which are poorly known. Details of the impact-ejecta model are described in \citet{Krivov:2003ku} and \citet{Spahn:2006dh}. The important parameters used here are listed in Table~\ref{tab:impeje}. The resulting source rates $\dot{N}^+$ are summarize in Table.~\ref{tab:n_plus}.

\begin{table*}[htpb]
    \centering
    \begin{tabular}{ccl}
        name & value & definition \\
        \hline
        $F_{imp}$ & $\sim 9 \times10^{-17} $ & [$\gm\cm^{-2}\s^{-1}$]; impactor flux, include gravitational focusing $\sim$5 \\
        $g_{sil}$ & 0 & ratio of silicon/ice on surface of the targets\\
        $K_e/K_i$  & 0.1 & ratio of ejecta's and impactor's kinetic energy \\
        $\gamma$ & 2.4 & cumulative mass distribution of ejecta
    \end{tabular}
    \caption{Parameters for impact-ejecta model \citep{Krivov:2003ku,Spahn:2006dh} with Methone and Anthe as targets with smooth icy surfaces.}
    \label{tab:impeje}
\end{table*}

\begin{table}[htpb]
    \centering
    \begin{tabular}{cccccc}
        & R ($\km$) & $Y$ & $v_{esc}$ (m/s) & $v_0$ (m/s) &  $\dot{N}^+$ ($s^{-1}$)  \\
        \hline
        Methone & 1.45  & 18232 & 0.49  & 2.65  & $1.7\times10^{8}$ \\
        Anthe   & 0.5   & 17922 & 0.14 & 2.66  & $2.0\times10^{7}$ \\
    \end{tabular}
    \caption{The mean radii of Methone and Anthe, the yield $Y$, esscape speed $v_{esc}$, initial ejecta speed $v_0$, and dust production rate $\dot{N^+}$ base on impact-ejecta model with parameters in Table.~\ref{tab:impeje}. To obtain $v_0$, we assumed $K_e/K_i = 0.1$ and the index of the power law ejecta speed distribution equals to 2. The mean radii are from \citet{Thomas:2013bz}. }
    \label{tab:n_plus}
\end{table}

\subsection{Dust dynamics}
\label{sec:dynamics}
The equation of motion for micron-sized particles in Saturn's E ring are given by
\begin{equation}
    \label{eq:motion}
    \begin{split}
        m \ddot{\mathbf{r}} = & -m \nabla\Phi_{S} -
        \sum_i G M_i m \frac{\mathbf{r}-\mathbf{r}_{i}}{\abs{\mathbf{r}-\mathbf{r}_{i}}^3} \\
        & + q \left[(\dot{\mathbf{r}} - (\Omega_s \times \mathbf{r})) \times \mathbf{B} \right] -
        \frac{I_{\odot} \sigma Q_{pr}}{c} \hat{e}_\odot + \mathbf{F}_{D}
    \end{split}
\end{equation}
with the particle mass $m$ and its acceleration $\ddot{\mathbf{r}}$.
The vector $\vec{r}$ measures the location of the dust in an inertial frame originating in Saturn's center.
On the right hand side, we have the gravity of oblate Saturn $\nabla\Phi_{S}$, where the gravitational potential $\Phi_{S}$ of an oblate body can be found in, for example, \citet{Murray:1999th}.
The second term is the gravity of the moons, where $M_i$ and $\mathbf{r}_i$ are mass and position of the $i$th moon and $G$ is the gravitational constant. Here we consider the gravity of Methone or Anthe, Mimas, and Enceladus simultaneously. The former two are the source moons of dust, Mimas introduces the resonances near Methone and Anthe orbits, and Enceladus is the other large perturbing moon nearby.
The third term is the Lorentz force ($\mathbf{F}_{L}$), caused by the relative motion between the particle and Saturn's magnetic field. The charge of particles are denoted by $q$ and the term $(\Omega_s \times \mathbf{r}) \times \mathbf{B}$ can be interpreted as a corotational electric field, with spin rate of Saturn $\Omega_s=1.6216\times10^{-4} \,\mathrm{rad}/\s$ and Saturn's magnetic field $\mathbf{B}$, which is assumed to be a aligned dipole with the magnetic field strength at Saturn's equator -0.21 Gauss. The charge is simply modeled by a constant surface potential of $-4$~V on a spherical dust grain. The relation of surface potential $\phi$ and charge $q$ is $\phi = (1/4 \pi \epsilon_0)(q/s)$ where $\epsilon_0$ is the vacuum permittivity and $s$ the radius of particle.
The solar radiation pressure ($\mathbf{F}_{\odot}$) is the fourth term, where $I_{\odot} \approx 14 \,\mathrm{W} / \mathrm{m}^2$ is the solar energy flux at Saturn,
$c$ is the speed of light,
$Q_{pr}$ is the radiation pressure efficiency which is of order of unity, and is about 0.3 for icy particles with radius several micrometer or larger \citep{Burns:1979bg}. Further,
$\sigma = \pi s^2$ denotes the cross section of the particle.
and $\hat{e}_\odot$ is a unit vector pointing from the particle toward the Sun.
The last term is the plasma drag $\mathbf{F}_{D}$ given by Eq.~\eqref{eq:Fd} later, which is dominated by the direct collisions between the particle and the bulk of plasma.

In the following, we briefly introduce the Solar radiation pressure, plasma drag, and a confinement of the arcs by Mimas resonances. 

\subsubsection{Solar radiation pressure}
\label{sub:solar_radiation_pressure}

Solar radiation pressure $\mathbf{F}_\odot$ leads to an increasing eccentricity, but due to the additional precession of the elliptical orbit due to Lorentz force and planetary oblateness the eccentricity performs a sinusoidal oscillation. A similar effect can be also observed in vertical direction for the inclination, although the passage of the particle through the planetary shadow is necessary.
The orbital elements perturbed by $\mathbf{F}_\odot$ are then given \citet{Hedman:2010ie}:
\begin{equation}
    e = \frac{n}{\dot{\varpi}_0'} \left[
          3 f(\epsilon)  \frac{F_{\odot}}{F_S} \cos{B_{\odot}}
        \right]
        \sin{(\dot{\varpi}_0 t / 2)}
    \label{eq:e}
\end{equation}
\begin{equation}
    \varpi - \lambda_\odot = \bmod \left(\frac{\dot{\varpi}_0 t}{2}, \pi\right) + \frac{\pi}{2}
    \label{eq:varpi}
\end{equation}
\begin{equation}
    i = 2 \frac{n}{\dot{\Omega}_0'} \left[
          g(\epsilon) \frac{F_{\odot}}{F_S} \sin{\abs{B_{\odot}}}
        \right]
        \sin{(\dot{\Omega}_0 t / 2)}
    \label{eq:i}
\end{equation}
\begin{equation}
    \Omega - \lambda_\odot = \bmod\left(\frac{\dot{\Omega}_0 t}{2}, \pi\right) + \pi  \frac{B_\odot}{\abs{B_\odot}} \,.
    \label{eq:Omega}
\end{equation}
where $e$ is eccentricity, $\varpi$ is longitude of pericenter, $i$ is inclination, $\Omega$ is longitude of ascending node, $n$ is the mean motion of particles, $B_\odot$ is the elevation angle of the Sun relative to ring plane (not to be confused with the Saturn's magnetic field $\mathbf{B}$ in Eq.~\eqref{eq:motion}), and $\lambda_\odot$ is the longitude of the Sun.
The precession rate of pericenter and precession rate of ascending nodes are $\dot{\varpi}_0'=\dot{\varpi}_0 - \dot{\lambda}_\odot$ and $\dot{\Omega}_0' = \dot{\Omega}_0 - \dot{\lambda}_\odot$, respectively, where $\dot{\varpi}_0$ and $\dot{\Omega}_0$ are precession rates introduced by the combination of planetary oblateness and Lorentz force, $\dot{\lambda}_\odot$ is the slow seasonal variation of the solar longitude. The precession rate by planetary oblateness is $\dot{\varpi}_{J2} \approx \abs{\dot{\Omega}_{J2}} \approx 0.85^\circ$/day in Methone orbit and 0.8$^\circ$/day in Anthe orbit, while by Lorentz force $\dot{\varpi}_L=-0.66(\phi/{-}4\,V)(1\mum/s)^2\,\mathrm{deg}/\mathrm{day}$ in Methone orbit and $\dot{\varpi}_L=-0.63(\phi/{-}4\,V)(1\mum/s)^2 \,\mathrm{deg}/\mathrm{day}$ in Anthe orbit.
In practice, $\dot{\lambda}_\odot$ is very small (${\sim}12^\circ/\yr \approx 0.03^\circ/day)$ in comparison to the precession rates dominated by planetary oblateness or Lorentz force and thus will be ignored for the current problem.
The definitions $f(\epsilon) = 1 - \epsilon + \sin(2\pi\epsilon)/6\pi$, and $g(\epsilon) = \sin(\pi\epsilon)/\pi$ are used in \citet{Hedman:2010ie} obtained by the averaging over one particle orbit, where $\epsilon$ is the fraction passage time of the shadow during one particle orbit.

By applying the model above, particles near Methone achieve a forced eccentricity and inclination
\begin{equation}
    e_{f} \approx 4.69\times10^{-2} f(\epsilon) \cos{B_\odot} \frac{Q_{pr}}{s/1\mum}
    \label{eq:M_e_max}
\end{equation}
\begin{equation}
    i_{f} \approx 3.13\times10^{-2} g(\epsilon) \sin{\abs{B_\odot}} \frac{Q_{pr}}{s/1\mum},
    \label{eq:M_i_max}
\end{equation}
and slightly different at Anthe orbit
\begin{equation}
    e_{f} \approx 4.38\times10^{-2} f(\epsilon) \cos{B_\odot} \frac{Q_{pr}}{s/1\mum}
    \label{eq:A_e_max}
\end{equation}
\begin{equation}
    i_{f} \approx 2.93\times10^{-2} g(\epsilon) \sin{\abs{B_\odot}} \frac{Q_{pr}}{s/1\mum}
    \label{eq:A_i_max}
\end{equation}
where $\epsilon \leq 0.10$ at Methone and Anthe orbit, and $B_\odot$ varies between $\pm 26.7^\circ$. Note that $\epsilon$ is also a function of $B_\odot$, and $\epsilon$ can be zero while the Sun is high above the ring plane. Therefore for particles near Methone and Anthe orbits $f(\epsilon) \cos (B_{\odot})$ ranges from 0.89 to 0.93, whereas $g(\epsilon) \sin (\abs{B_{\odot}})$ spans values between 0 and 0.044.

\subsubsection{Plasma drag}
\label{sub:plasma_drag}
To estimate the strength of plasma drag one first needs to know the ion densities.
The sources of plasma in the innermost region of the E ring (3-4$\Rs$) are Enceladus and the ionization of the main ring $\rm O_2$ atmosphere \citep{Johnson:2006fs,Tseng:2010jb,Elrod:2012kb,Elrod:2014fe}. In this region, observations by Cassini spacecraft shows that the heavy ion densities (water group ions and $\Opp$) are about 1-100$\cm^{-3}$, and are highest near the Enceladus orbit \citep{Elrod:2014fe}. Furthermore, there are seasonal variations of the ion densities, i.e., the ion densities are higher while the Sun is high above ring plane \citep{Tseng:2010jb,Elrod:2012kb,Elrod:2014fe}. To simplify the problem, we assume there are $\Op$ ions with constant densities of $43.2\cm^{-3}$ in Methone's orbit and $44.0\cm^{-3}$ in Anthe's orbit.

The force due to plasma direct collision is given by \citep{Banaszkiewicz:1994kp}
\begin{equation}
  \begin{split}
    \label{eq:Fd}
    F_{D} = \pi n_i m_i u_i^2 s^2
        & \bigg[  (M_i + \frac{1}{2 M_i}) \frac{\exp(-M_i^2)}{\sqrt{\pi}} + \\
        & (M_i^2 + 1 - \frac{1}{4 M_i^2}) \erf(M_i) \bigg].
  \end{split}
\end{equation}
The ions are characterized by their number density $n_i$, thermal velocity $u_i$, and mass $m_i$, respectively. The Mach number is defined by $M_i = \tilde{v}_0/u_i$, where $\tilde{v}_0$ is the velocity of a particle relative to bulk of plasma. The Mach number relative to $\Op$ ions in Methone and Anthe orbit are both about 1.2.

Plasma can also interact with charged particles via Coulomb forces \citep{Northrop:1990jx}. This indirect `Coulomb drag' is ignored in our model, because at 3-4$\Rs$ the Coulomb drag is much weaker than the contribution via plasma direct collisions $\mathbf{F}_D$. In the Saturn environment, Coulomb drag is important only where the Mach number is about unity, e.g. in the A ring, including the Encke gap (\citealt{Grun:1984wx}; Sun 2015, in preparation).

In the inner part of the E ring, plasma nearly corotates with Saturn. The corotation speed is faster than the Keplerian speed of particles outside the synchronous radius of 1.86$\Rs$. Collisions with ions push particles roughly in direction of the orbital motion, this means particles gain orbital energy and angular momentum and the particle is migrating outward. By utilizing the Gaussian perturbation equations \citep[e.g.,][]{Burns1976}, the plasma drag induces an increase of semi-major axes near Methone and Anthe in the rate
\begin{equation}
    \label{eq:dadt}
    \frac{da}{dt} \approx 600 \left(\frac{1\mum}{s} \right) \km \yr^{-1}.
\end{equation}
In other words, 1$\mum$ particles can migrate outward by about 0.3$\Rs$ in one Saturnian year.

\subsubsection{Mimas resonances and perturbing forces}
\label{sub:mimas_resonances}
Particles ejected from the surface of Methone and Anthe are likely being trapped in the arcs, same as their source moons. However perturbing forces complicates the situation, especially the drag force cause particles to drift out of the resonance. Roughly speaking, the probability of trapping migrating particles is higher if the migration rate is lower \citep[e.g.,][]{Dermott:1994gb,Wyatt:2003fp,Vitense:2012eu,Shannon:2015cg}.
In this case, the migration rate $da/dt$ is induced by plasma drag and the value is inverse proportional to particle size (Eq.~\eqref{eq:dadt}), this means large particles migrate slower and therefore are more likely being trapped in resonances.

\subsection{Sinks}
\label{sub:sinks}
Particles ejected from Methone and Anthe may get caught in the arcs or escape then continuously migrate outward due to plasma drag. For particles in the arcs, the dominating particle sinks are the parent moons embedded in the arcs. Particles escaping the arcs migrate outward until they reach Enceladus' orbit, where they collide with the moon or drift beyond it's orbit.

There exists an upper limit of lifetime dominated by plasma sputtering \citep{Johnson:2008jz}. The plasma sputtering rate in the inner E ring is in the order of 1$\mum$ radius every 100 years \citep{Johnson:2008jz}. We use a slightly smaller value (0.6$\mum$ per 100 years) because the sputtering rate at Methone and Anthe orbits is slightly smaller because of the lower plasma density \citep[see Fig.~3 in ][]{Johnson:2008jz}. With this value, the sputtering lifetime is
\begin{equation}
    T_{sput} = 167 \left(\frac{s}{1\mum}\right) \,\mathrm{years}
    \label{eq:t_sput}
\end{equation}
where $s$ is the particle radius. This sputtering lifetime defines the maximal possible particle lifetime.
They never reach such long lifetimes. When the particle radius drops below a critical value, the particle escapes the resonance and migrates radially outwards and leaving the zone of interest.

In general, the erosion of particles means a change in the particle dynamics with time. Namely, $e$ and $i$ and $da/dt$ are all increasing during erosion (Eq.~\eqref{eq:M_e_max}, Eq.~\eqref{eq:A_e_max}, and Eq.~\eqref{eq:dadt}). Beside particles gain probability to escape the resonance, they also migrate faster, and gain larger eccentricities and inclinations as time goes on.

\section{Numerical Simulations}
\label{sec:results}

\subsection{Configuration of simulation}
In order to find out the distribution of dust near Methone and Anthe, the equations of motion Eq.~\eqref{eq:motion} have been numerically integrated using the RADAU integrator \citep{Everhart:1985vu} implemented in the MERCURY package \citep{Chambers:1999jd}.
In addition to the perturbing forces, the plasma sputtering rate of 0.6$\mum$ per 100 year is considered, i.e., particle radii are decreasing with time.
The simulation stops once all test particles are removed -- either by collision with Methone, Anthe, Mimas, Enceladus, or when their radial position is inside Mimas orbit or outside Enceladus orbit.

Particles are ejected randomly from the surface of Methone and Anthe with speed 2.6~$\m\s^{-1}$ (the initial ejecta speed as given in Table~\ref{tab:n_plus}) and constant elevation angle of 45$^\circ$ relative to surface of moon. The speed is larger than the escape speed of both moons, which means almost all ejecta can escape. The radius of the ejecta $s$ is assumed to obey a power law distribution $n(s) \sim s^{-\gamma}$, with $\gamma=2.4$.

Instead of using the power law distribution directly, we choose the particle radii randomly between 1 and 30$\mum$ in the simulation and weight the resulting density with the according power law distribution.
The benefit of weighting later is that one do not need to run simulations with a lot of tiny particles and only a few large particles, which means one need to huge amount of particles to reach reaonsable numerical result. Similar procedures has been applied to Phobe ring \citep{Hamilton:2015kh} in comparing the size distribution and observation.

We use the following procedure to map the simulation data to the real numbers. After weighting to the power law size distribution, the total number of particles in the beginning of simulation is $N_{sim0}$, this number should be scaled to number of dust produced per unit time, $\dot{N}^{+}$.
In the simulation, we only eject particle at one single time and have particle orbits stored in equal time steps $\Delta t$. To map the simulation data to real numbers, we count the cumulative number of particles in a `box' $N_{sim}$, the box can be a grid in $x-y$ plane or in longitude-radius grids.
After all these procedure, the real number of particles in the box is
\begin{equation}
    \label{eq:scale_to_real}
    N = N_{sim} \frac{\dot{N}^+}{N_{sim0}} \Delta t.
\end{equation}

One further thing to notice is that both Methone and Anthe are both librating inside the corresponding arcs, meaning the ejecta are ejected while the moon is at different phase of the librations. The initial condition of the moon, which is similar to that of ejecta, should have impact on the amount of time for ejecta to stay in the arcs. Therefore in the simulation particles are ejected while the source moons are at different resonant arguments of the Mimas resonances.

\subsection{Methone and Anthe arcs}
\label{sub:arcs}
\begin{figure}[htpb]
    \centering
    \includegraphics[width=0.97\linewidth]{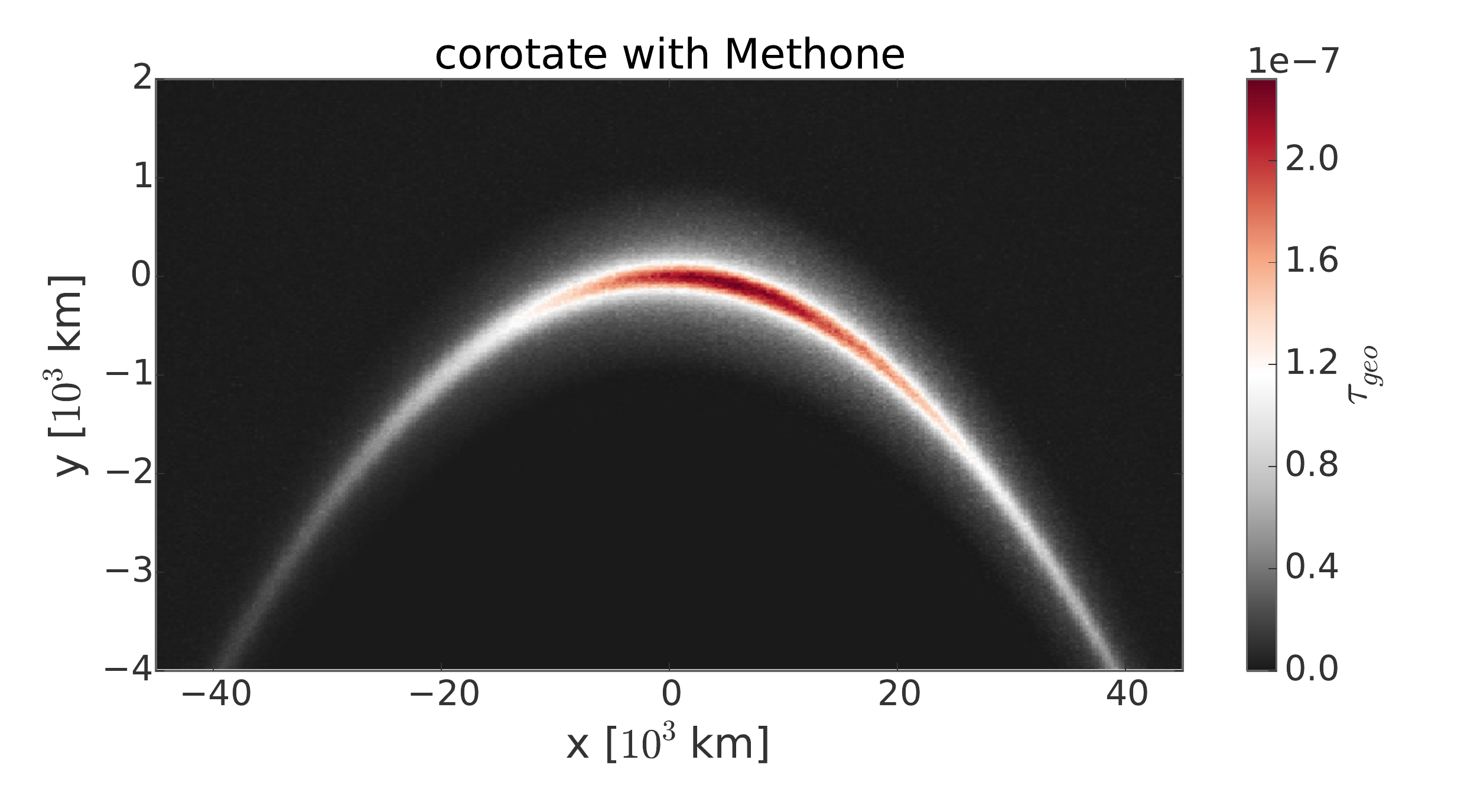}
    \caption{Simulation results giving the geometric optical depth graph in the corotation frame of Methone. Methone is at origin and orbiting Saturn counterclockwise (toward $-x$). Saturn is at direction of $-y$. This figure covers about 24$^\circ$ of Methone orbit.}
    \label{fig:m_corot}
\end{figure}

\begin{figure}[htpb]
    \centering
    \includegraphics[width=0.97\linewidth]{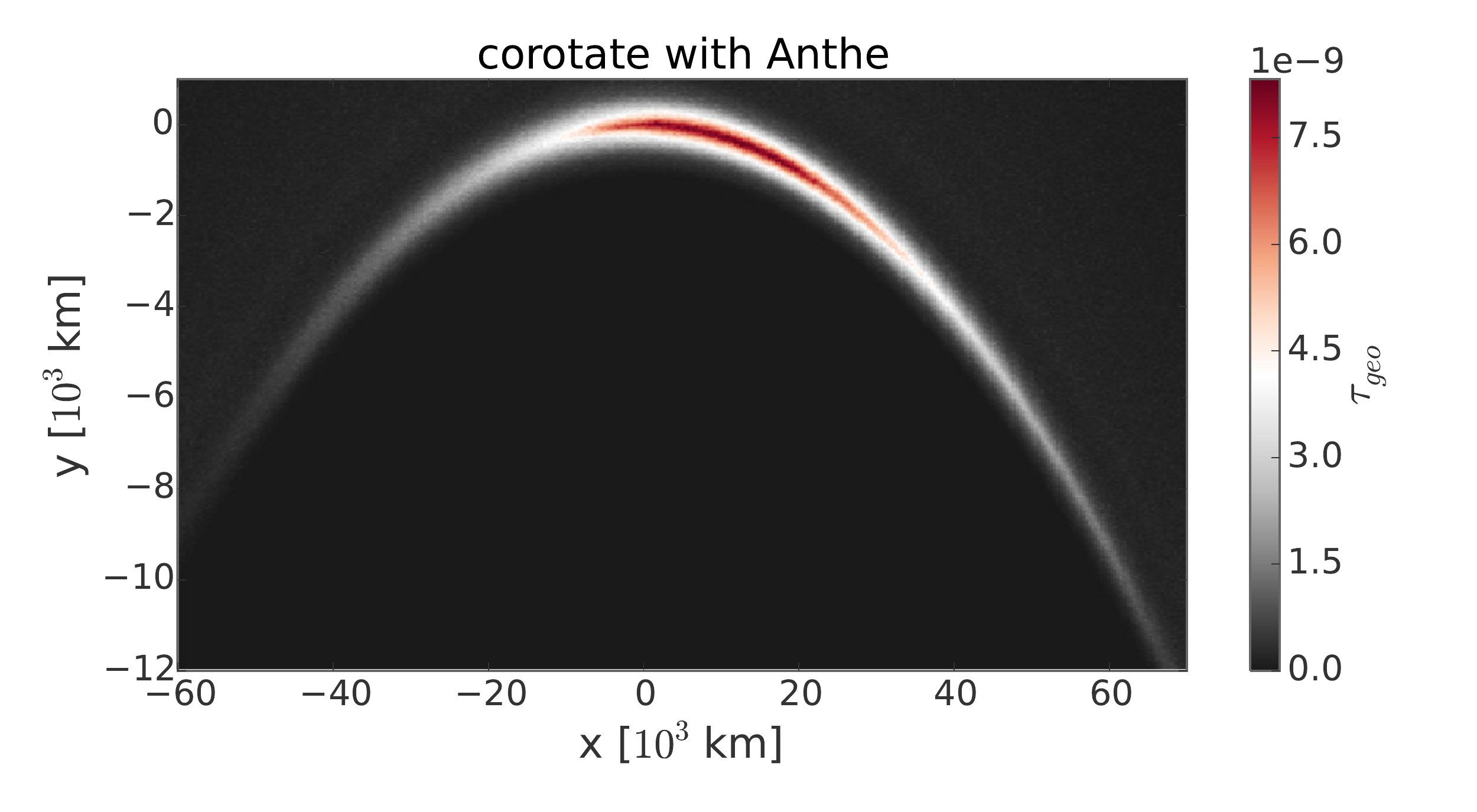}
    \caption{Simulation results as a geometric optical depths graph in the corotation frame of Anthe. Anthe is at origin and orbiting Saturn counterclockwise (toward $-x$) and Saturn is at direction of $-y$. This figure covers about 36$^\circ$ of Anthe orbit.}
    \label{fig:a_corot}
\end{figure}

\begin{figure}[htpb]
    \centering
    \includegraphics[width=0.97\linewidth]{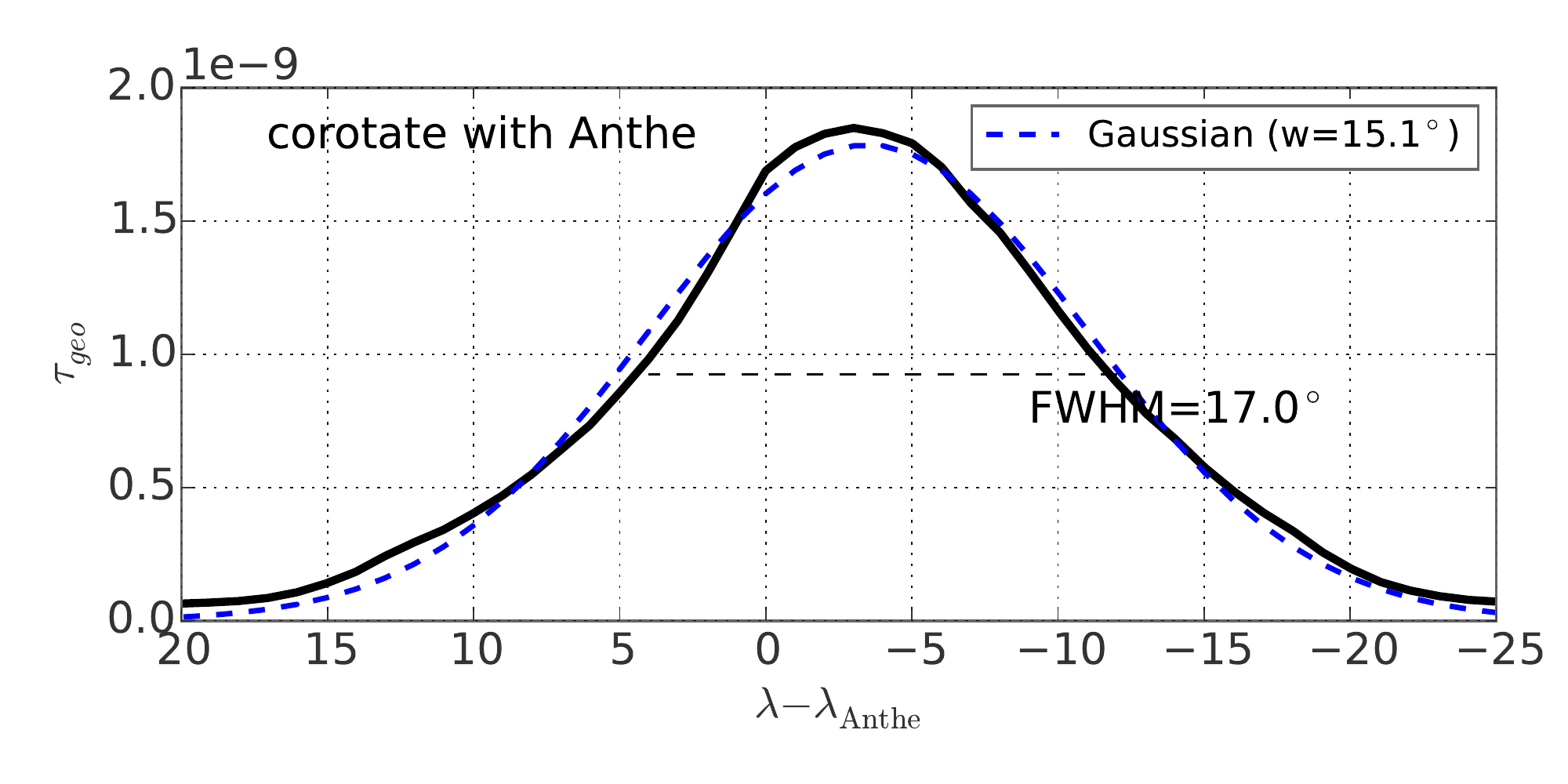}
    \caption{Longitudinal scan in corotational frame of Anthe. The optical depth is smaller than Figure~\ref{fig:a_corot} is because here use larger box. The blue dashed line is the best fit of Gaussian function with width $\approx15.1^\circ$, while the full width at half maximum (FWHM) is about 15.1$^\circ$.}
    \label{fig:a_lme_scan}
\end{figure}

We have simulated the entire `life' of about 2000 test particles starting at both Methone and Anthe. Fig.~\ref{fig:m_corot} and Fig.~\ref{fig:a_corot} show the geometric optical depth in the corotating frame of Methone and Anthe. Geometric optical depth is the optical depth without considering any of the light scattering by particles, in other words, it's the ratio of the total cross section of ring particles in a unit area to the unit area.
The source moons are in the origin of both figures and orbiting Saturn counterclockwise. With the simulation we have found an ${\sim}20^\circ$ arc near Methone and ${\sim} 30^\circ$ arc in Anthe orbit.
Nevertheless, in reality the arcs are embedded in the E ring background, making it difficult to observe the full length of the arcs, especially the low density extensions. Thus the full width at half maximum (FWHM) might be a better measurement for comparison with observation. In this case we get FWHM ${\sim}12^\circ$  for Methone's arc (see later in Fig.~\ref{fig:m_corot_sym_1D}) and ${\sim}17^\circ$ for Anthe's arc (shown in Fig.~\ref{fig:a_lme_scan}), while observations by \citet{Hedman:2009kt} yielded about $10^\circ$ and $20^\circ$, respectively.

\begin{figure}[htpb]
    \begin{subfigure}
        \centering
        \includegraphics[width=0.48\linewidth]{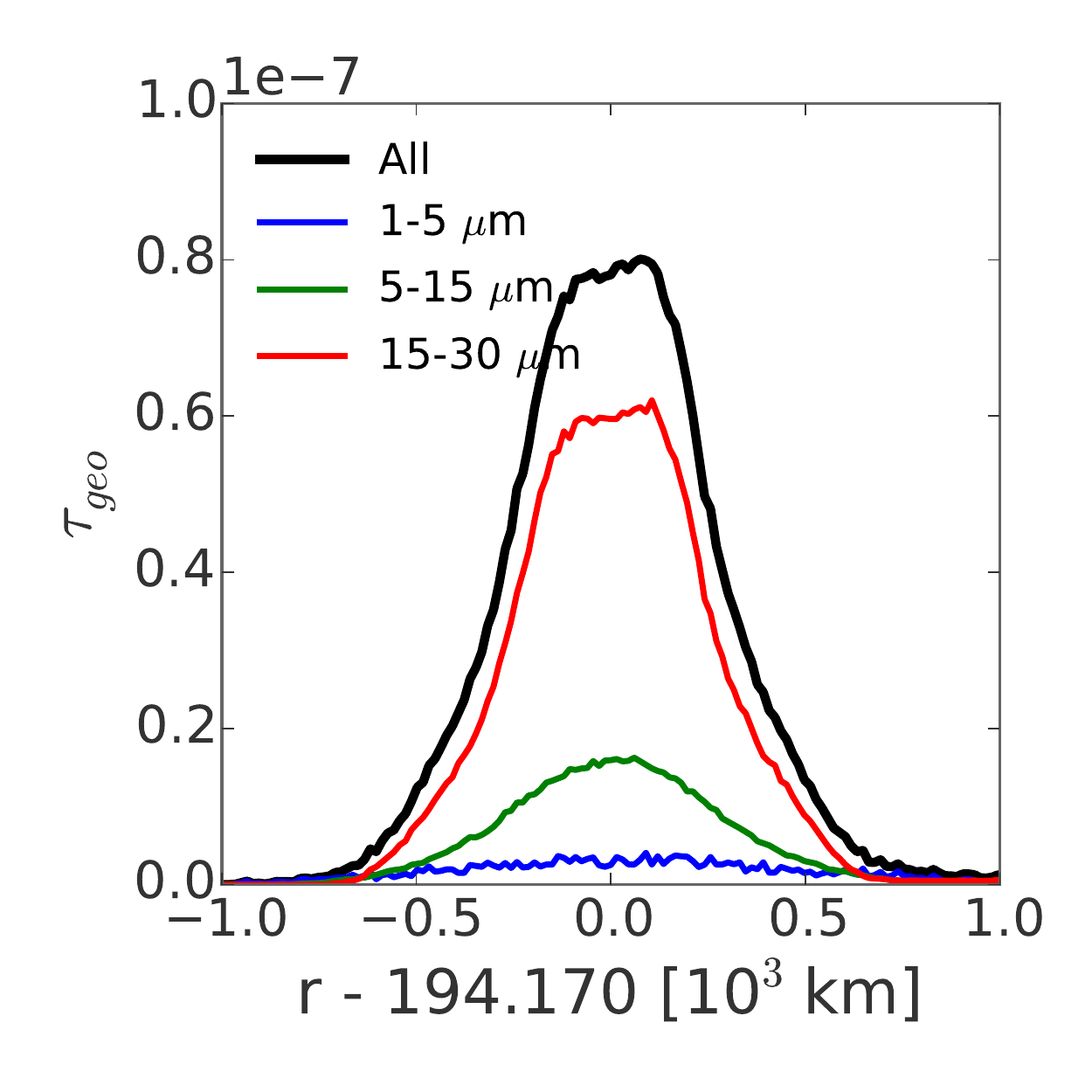}
    \end{subfigure}
    \begin{subfigure}
        \centering
        \includegraphics[width=0.46\linewidth]{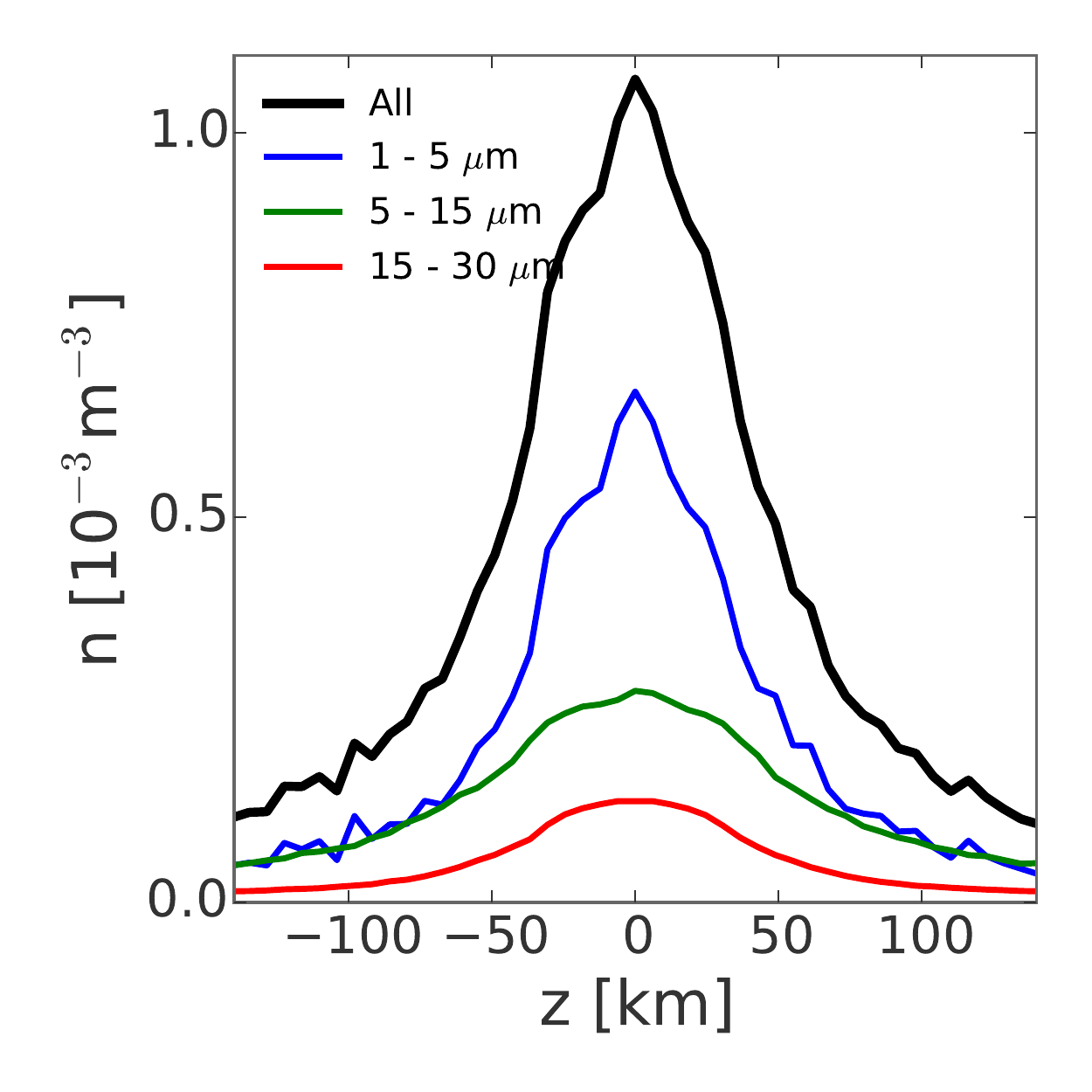}
    \end{subfigure}
    \caption{vertical and radial distribution of particles in Methone's arc. In the arc, the number of smaller particles is several times more than the larger ones (the right panel), but the total geometric optical depth is dominated by the larger ones (the left panel).}
    \label{fig:m_rz_hist}
\end{figure}

\begin{figure}[htpb]
    \begin{subfigure}
        \centering
        \includegraphics[width=0.48\linewidth]{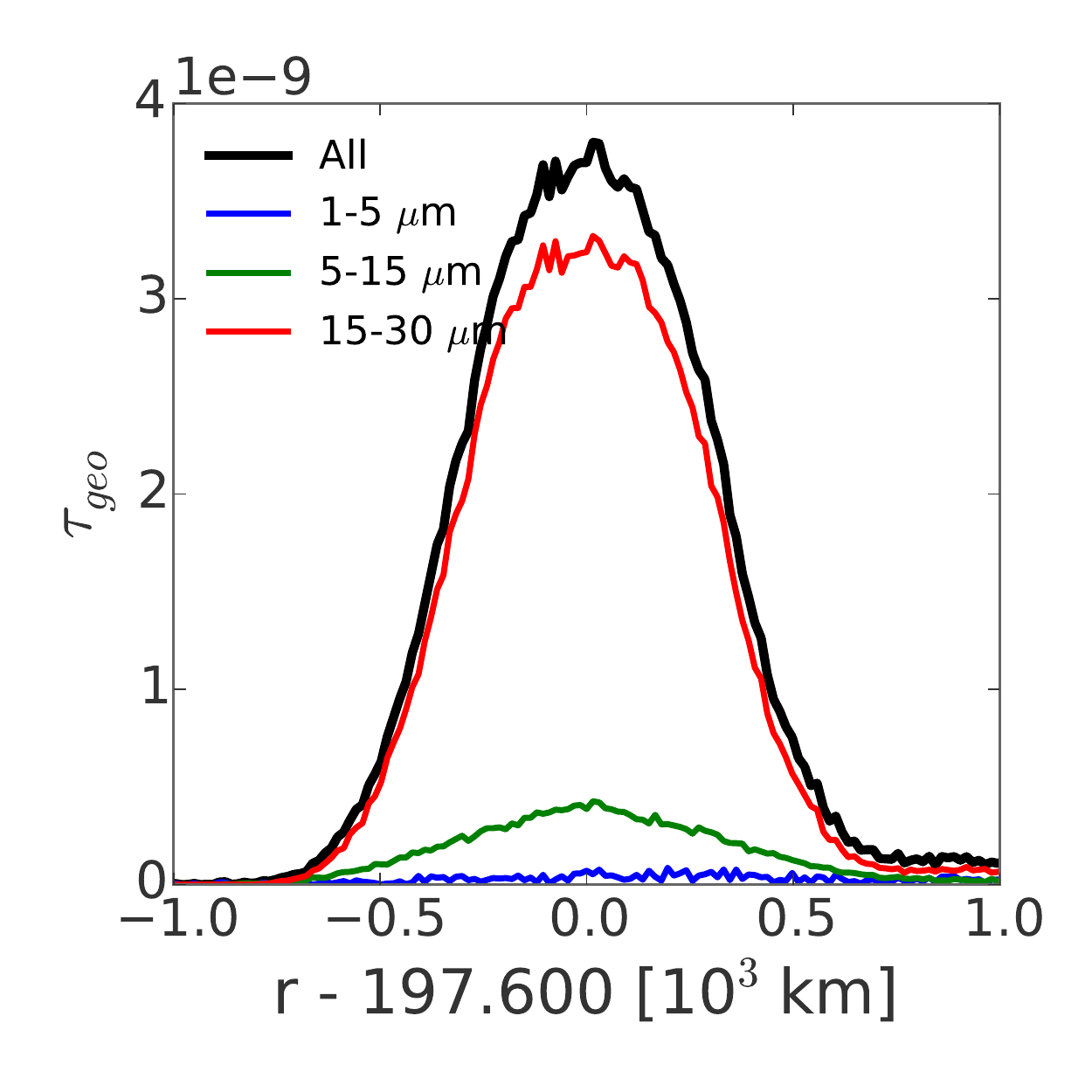}
    \end{subfigure}
    \begin{subfigure}
        \centering
        \includegraphics[width=0.46\linewidth]{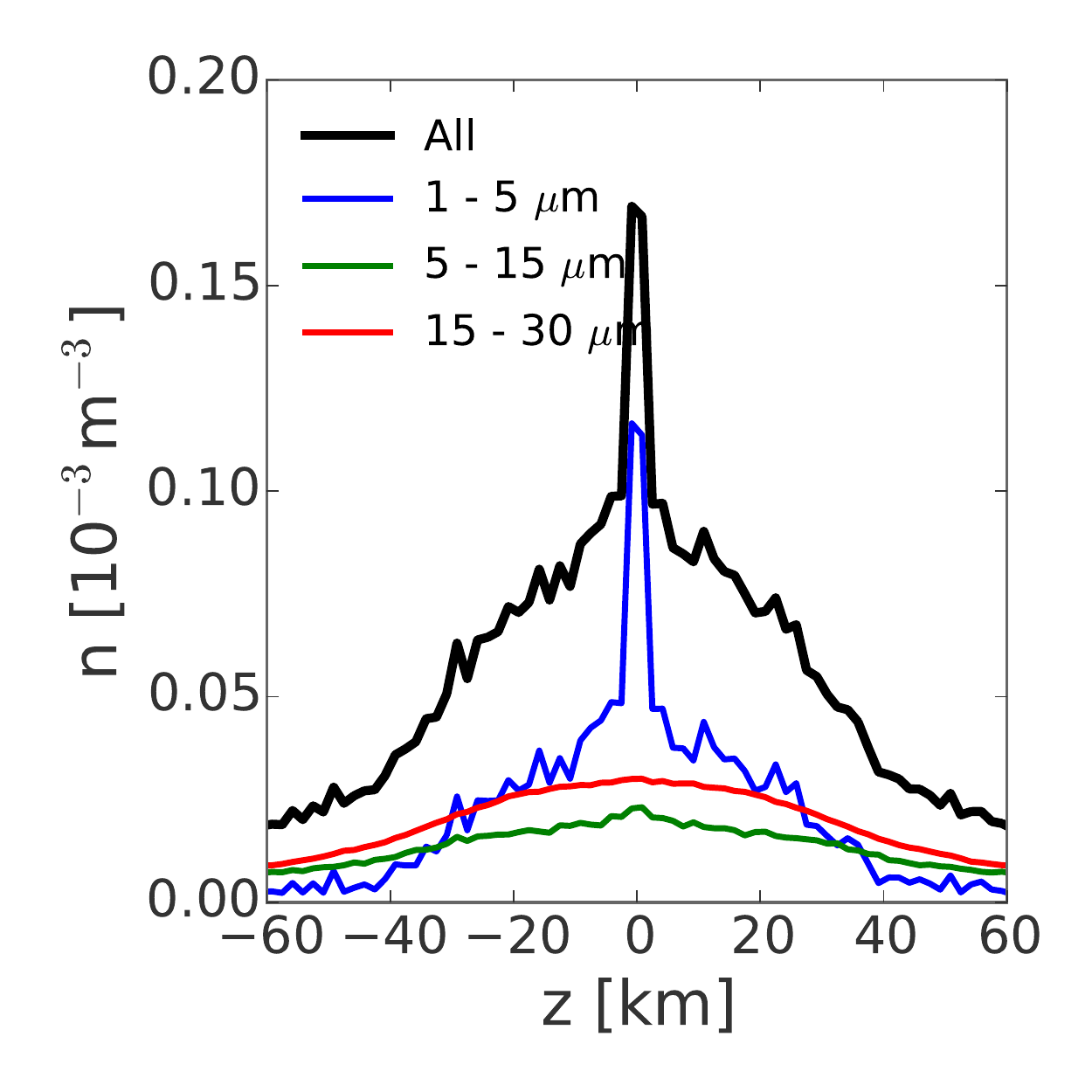}
    \end{subfigure}
    \caption{vertical and radial distribution of particles in Anthe's arc. The amount of large particles and smaller ones are about the same order (right panel), but the total cross section is dominated by the larger ones (left panel).}
    \label{fig:a_rz_hist}
\end{figure}

The radial profiles of the arcs are presented in the left panels of Fig.~\ref{fig:m_rz_hist} and Fig.~\ref{fig:a_rz_hist}. The radial extension of both arcs ($\approx 1000\km$ with FWHM$\approx 500 \km$) is dominated by the eccentricity introduced by solar radiation pressure and partly contributed by the eccentricities of the source moons, see Section~\ref{sub:heliotropic} for more detail.

The vertical distribution of the arcs are shown in the right panels of Fig.~\ref{fig:m_rz_hist} and Fig.~\ref{fig:a_rz_hist}. How are these inclinations generated? The inclinations of Methone and Anthe are 0.025$^\circ$ and 0.02$^\circ$, implying a vertical displacements ($a \times i$) of 84$\km$ and 69$\km$, respectively. These inclinations are also transferred to the trajectories of particles. On the other hand, the maximal forced inclinations introduced by solar radiation pressure (Eq.~\eqref{eq:M_i_max} and Eq.~\eqref{eq:A_i_max}) are about $(0.024/(s/1\mum))^\circ$.
Therefore, the vertical profiles for the particles in the arcs can be explained by a combination of source moon inclination and solar radiation pressure. The small particles in the Anthe arc should have larger inclinations due to solar radiation pressure but it is not seen here, indicating their early escape of the arc.

\begin{figure}[htpb]
    \centering
    \begin{subfigure}
        \centering
        \includegraphics[width=0.48\linewidth]{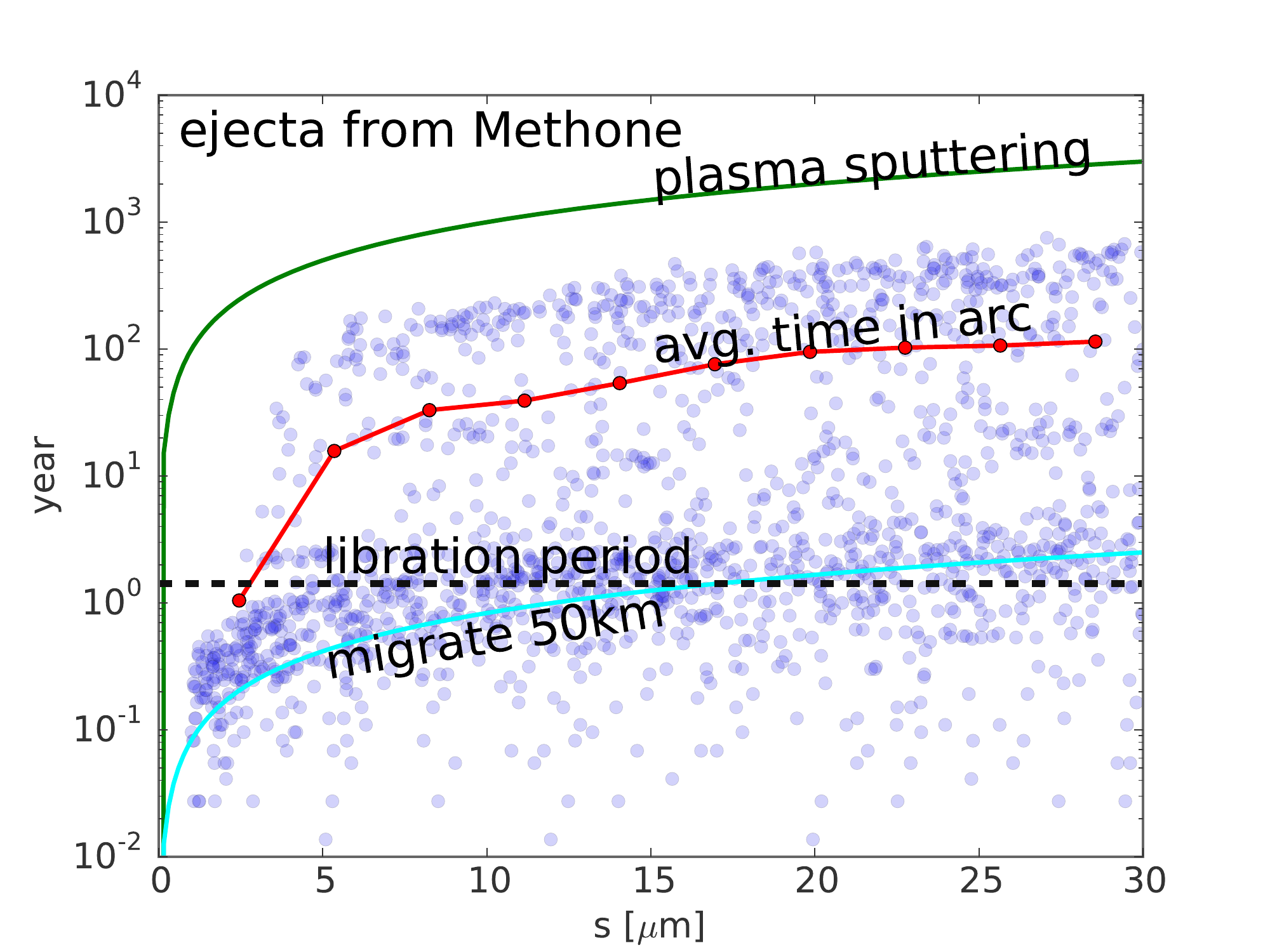}
    \end{subfigure}
    \begin{subfigure}
        \centering
        \includegraphics[width=0.48\linewidth]{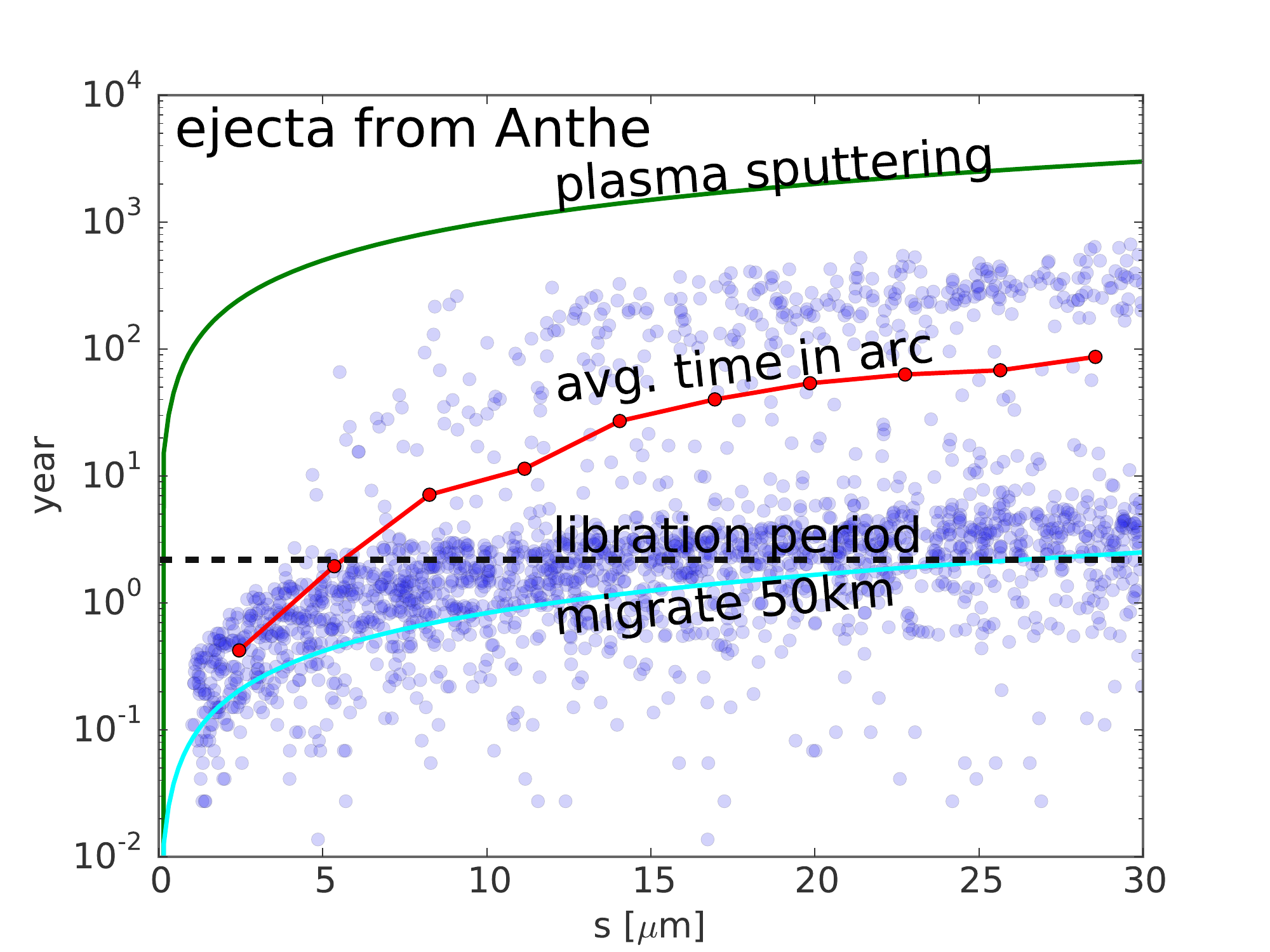}
    \end{subfigure}
    \caption{The relationship of initial particle radii and time spend in the arc. Each dot represent one particle. The green line is the sputtering lifetime $T_{sput}$ described in Eq.~\eqref{eq:t_sput}. The libration period is the time for the source moons to finish one libration around the stable co-rotation point, which is about 520 days for Methone and 800 days for Anthe. The cyan lines plots the time required for grain of radius $s$ to migrate outward by 50$\km$ if ther is only gravity from Saturn and plasma drag.}
    \label{fig:leave_arc_t0}
\end{figure}

It is expected that larger particles can stay in the arc for longer time due to smaller migration rate $da/dt$. We crudely estimate the time particles stay in arc by checking the time they reach certain semi-major axes. Simulations show that the librating amplitudes in radial direction for particles in the arcs are about $<40\km$, and we assume particles reaching 50$\km$ amplitude have left the arcs. This is not quite accurate but sufficient to see the trend.
The result is presented in Fig.~\ref{fig:leave_arc_t0}, which shows linear trends of maximal time particle stay in arcs and particle size.

\subsection{Heliotropic particles}
\label{sub:heliotropic}
\begin{figure}[htpb]
    \centering
    \includegraphics[width=0.97\linewidth]{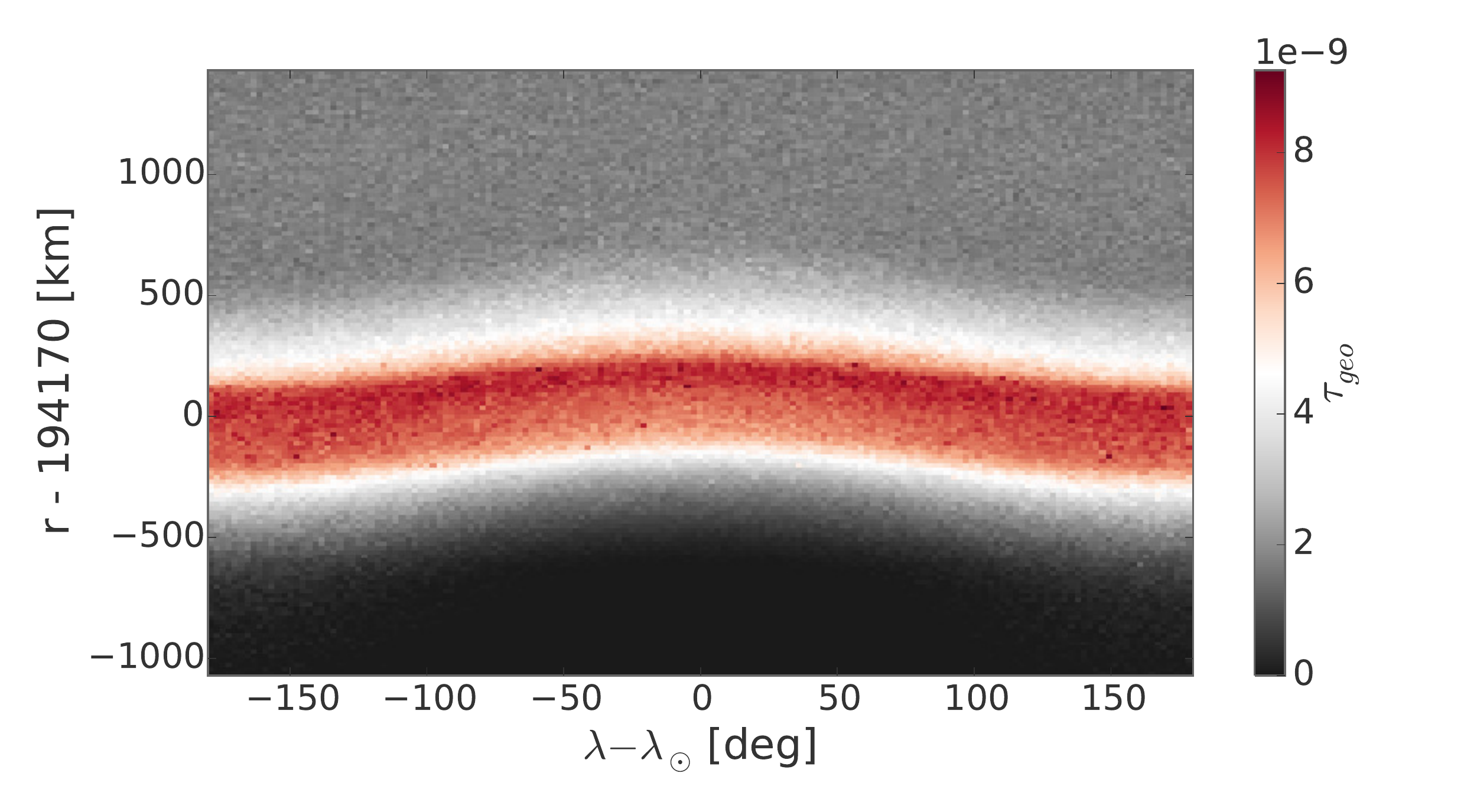}
    \caption{Heliotropic Methone `arc', accumulated over time and therefore spanning over all longitudes. $\lambda_\odot$ is longitude of the Sun and $\lambda$ is longitude of particle. The peak optical depth is smaller than the values in Fig.~\ref{fig:m_corot} or Fig.~\ref{fig:m_rz_hist}, because here considers the whole longitudes instead of only the arc region.}
    \label{fig:m_lme_r}
\end{figure}

Either in the arc or not, micron-sized grains are in heliotropic orbits, as shown in the simulation result in Fig.~\ref{fig:m_lme_r}. The \emph{average} pericenter of ring particles is oriented in anti-solar direction, same as predicted in Eq.~\eqref{eq:varpi}. The radial difference of pericenter and apocenter is about 1000$\km$, so eccentricity is about 0.0026. For comparison, the maximum eccentricity of Methone/Anthe are about 0.0015/0.001. Base on Eq.~\eqref{eq:M_e_max}, $e=0.0026$ implies a particle radius of $s = 5.4 f(\epsilon) \cos(B_\odot)$. In other words, the effective particle size is around 5$\mum$.

\subsection{Sinks of ejecta}
The sinks of all particles are summarized in Table~\ref{tab:fate}.
For ejecta from Methone, 0.1\% recollide to Methone, 18.1\% collide with Enceladus, and other 81.8\% reach orbits further outside. The fate of Anthe's ejecta are similar: 0.002\% recollide to Anthe, 19.3\% collide with Enceladus, and other 80.7\% escaped. Escaped particles either reach orbits outside the radial position of Enceladus or inside orbit of Mimas, either by large eccentricities induced by solar radiation pressure or outward migration by plasma drag.

\begin{table}[htpb]
    \centering
    \begin{tabular}{c|cccc}
        \multirow{2}{*}{source moon} & \multicolumn{4}{c}{sinks (\%)} \\
                & Methone &  Anthe & Enceladus & escaped \\
        \hline
        Methone & 0.1 &  -   & 18.1 & 81.8 \\
        Anthe   & -    & 0.002 & 19.3 & 80.7 \\
    \end{tabular}
    \caption{Sinks of arc particles in simulation. `Escaped` particles are removed from simulation after their radial distance is outside Enceladus orbit or inside Mimas orbit.}
    \label{tab:fate}
\end{table}

\begin{figure}[htpb]
    \centering
    \begin{subfigure}
        \centering
        \includegraphics[width=0.48\linewidth]{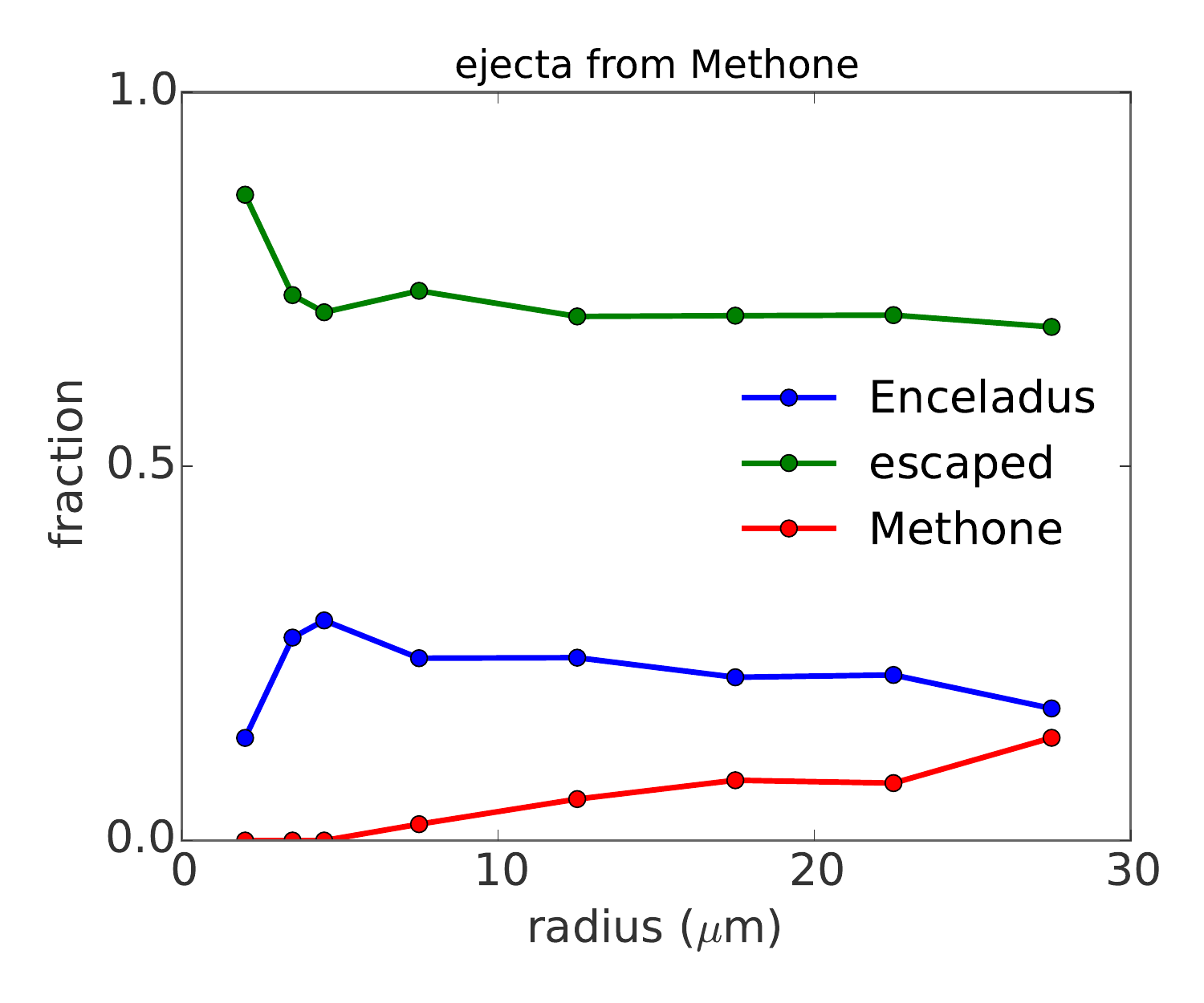}
    \end{subfigure}
    \begin{subfigure}
        \centering
        \includegraphics[width=0.48\linewidth]{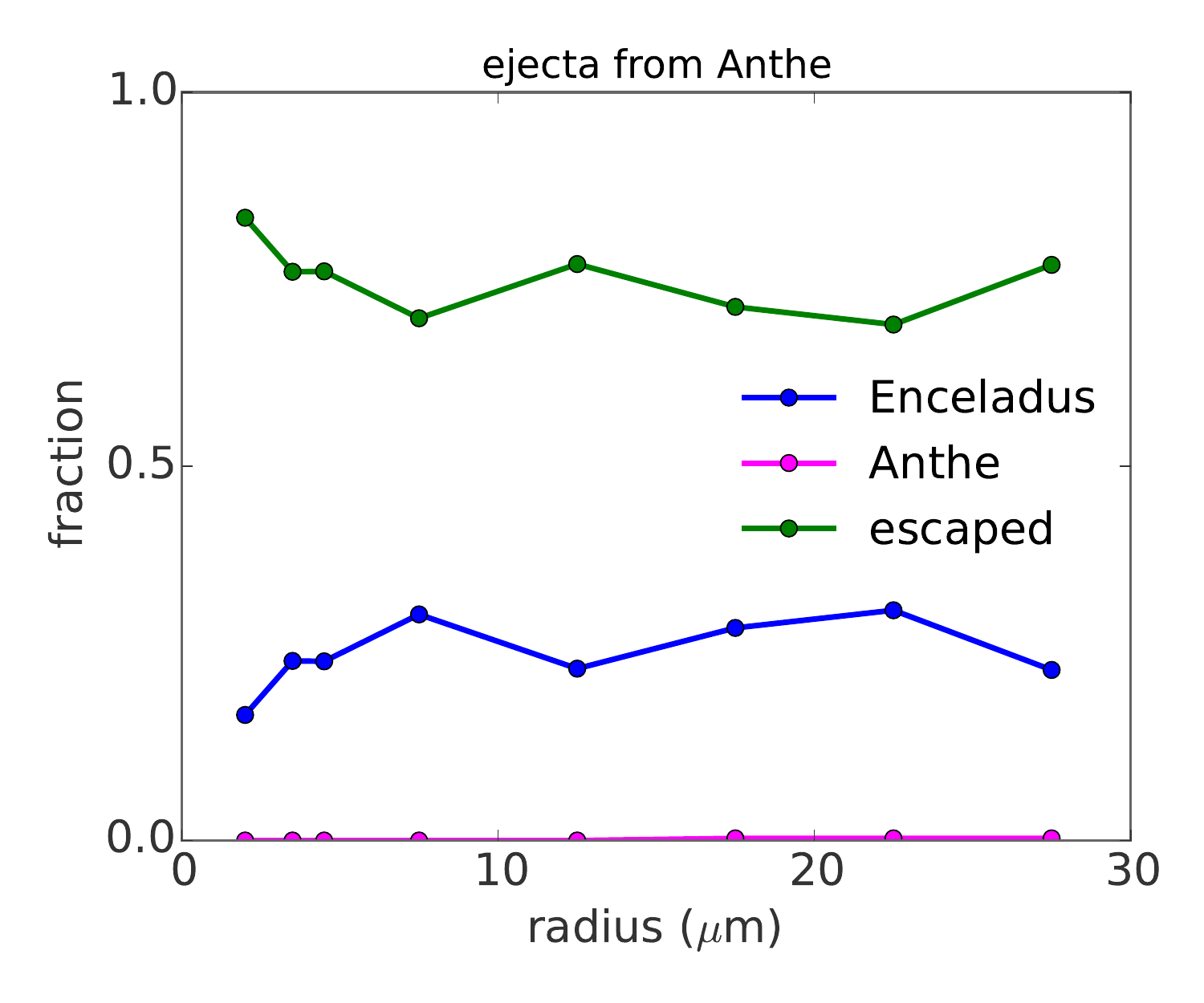}
    \end{subfigure}
    \caption{Percentage of sinks of particles by size. Left panel is for ejecta from Methone, right panel is for ejecta from Anthe.}
    \label{tab:size_fate}
\end{figure}

Due to the difference in their dynamics, large and small particles have quite different fates, as shown in Fig.~\ref{tab:size_fate}. Roughly speaking, larger particles from Methone are more likely to collide with the source moon because in average they are confined in the arcs close to the source for longer time. In the case of Anthe, probably due to the small cross section of Anthe, only a few large particles (out of 2000) collide with Anthe. Apart from that, for both Methone and Anthe, there are about 10-20\% of particles going to collide with Enceladus. The rest $\sim$80\% of particles they leave the area of interest (3-4$R_s$), due to outward migration of large eccentricity.

\section{Discussion and Conclusion}

\subsection{Simulation vs. Observation}
\begin{figure}[htbp]
    \centering
    \includegraphics[width=0.80\linewidth]{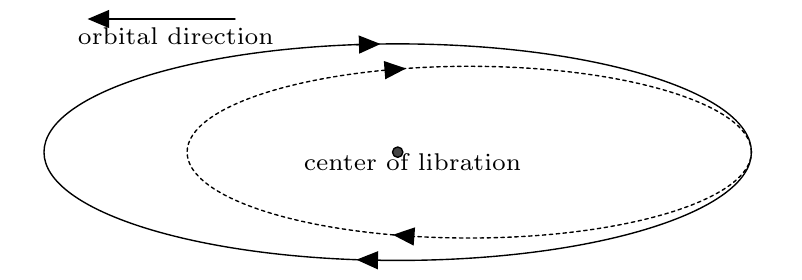}
    \caption{This cartoon illustrates how drag force cause the longitudinal asymmetry of arc, as seen in Fig.~\ref{fig:m_corot} and \ref{fig:a_corot}. Arc material are librating around center of libration. Without drag, particles librate inside the solid curve which correspond to the result in Fig.~\ref{fig:m_corot_sym}; with drag, particles librate inside the asymmetric dashed ellipse.}
    \label{fig:cartoon}
\end{figure}

\begin{figure}[htpb]
    \centering
    \includegraphics[width=0.93\linewidth]{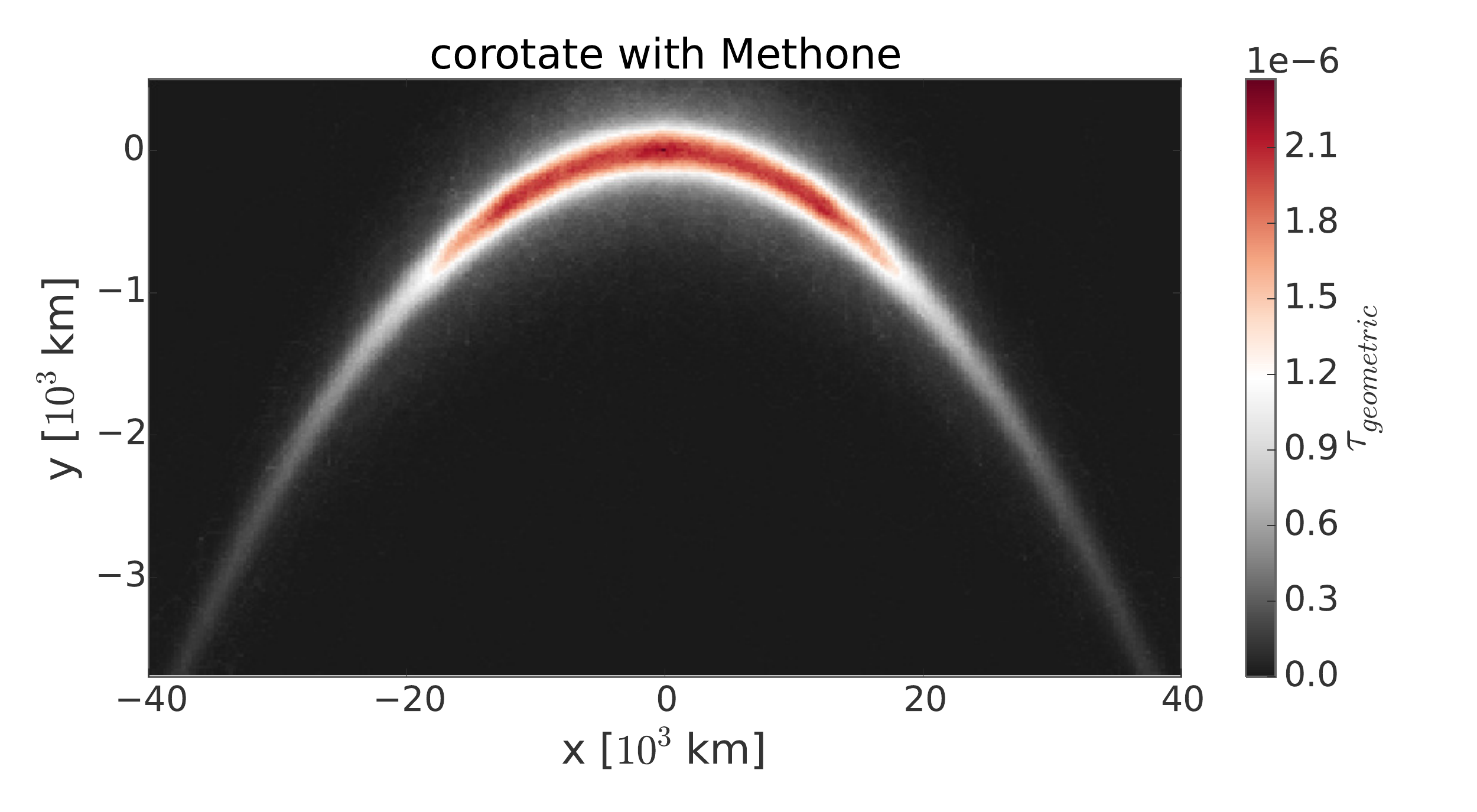}
    \caption{Similar to Figure~\ref{fig:m_corot} but plasma drag is not in the simulation and use only 160 test particles.}
    \label{fig:m_corot_sym}
\end{figure}

\begin{figure}[htpb]
    \centering
    \includegraphics[width=0.90\linewidth]{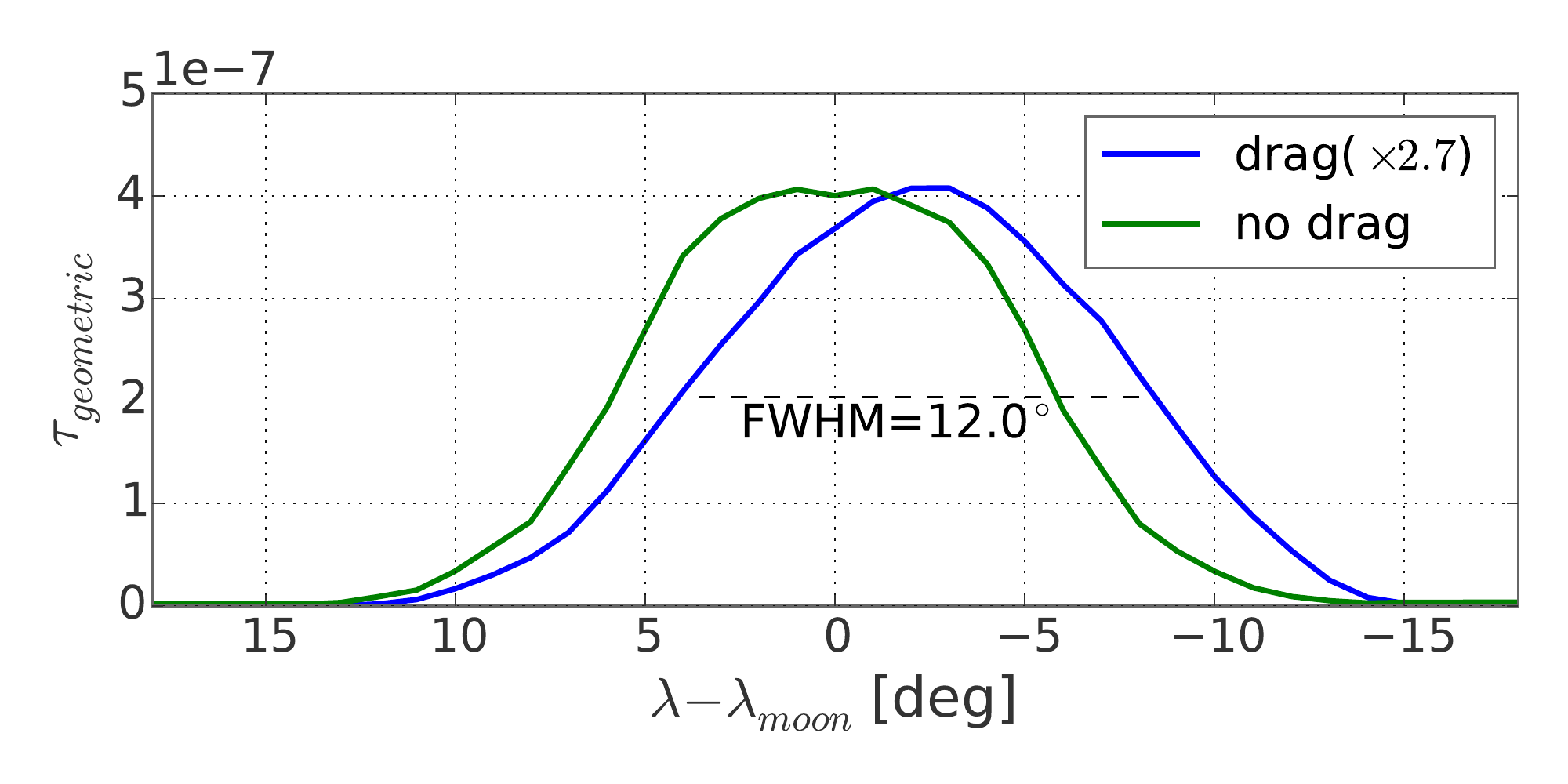}
    \caption{Longitudinal scan of Fig.~\ref{fig:m_corot} (drag) and Fig.~\ref{fig:m_corot_sym} (no drag). The maximal optical depth has been scaled to same level. The optical depths here are slightly smaller because the $\tau$ is averaged over a wider area (radial distance $194170\pm100\km$). The x-axis has been inverted so that the orbital direction is toward left, same as Fig.~\ref{fig:m_corot} and Fig.~\ref{fig:m_corot_sym} }
    \label{fig:m_corot_sym_1D}
\end{figure}

In the limited amount of observations by Cassini ISS, Anthe is found in the back (relative to orbital direction) and middle of the arc, this can be explained by the moon and arc material librate around a stable co-rotation point of the Mimas resonance at different phases \citep{Hedman:2009kt}. Our simulation results show additionally that the averaged relative longitudes of arc materials are behind Anthe's and Methone's mean position by several degrees, as shown in Fig.~\ref{fig:m_corot} and Fig.~\ref{fig:a_corot}.
This asymmetry can be explained by plasma drag as shown in Fig.~\ref{fig:cartoon}: while particles are librating around the certain stable co-rotation points of Mimas resonances, plasma drag prevent them to reach the furthermost region in the leading longitudes because they are pushed outward before they arrive at the region. Pushed outward means smaller Kepler velocity and therefore start to move backward relative to the center prematurely.
To verify this, we run a simulation of Methone arc without plasma drag. The result in Fig.~\ref{fig:m_corot_sym} shows a symmetric distribution of particle longitudes. While the longitudinal profiles of simulated Methone arc with and without plasma drag (Fig.~\ref{fig:m_corot} and Fig.~\ref{fig:m_corot_sym}) are shown in Fig.~\ref{fig:m_corot_sym_1D}. Note that such phenomenon has not been found in observations of Methone and Anthe arcs.
The same idea of the combination of drag force and resonance has been used to explain the longitudinal asymmetry of Encke ringlet of Saturn (\citealt{Hedman:2013hc}; Sun 2015, in preparation).

For the assumed values in the impact-ejecta model, we predict that the geometric optical depth should be in the order of $10^{-7} - 10^{-8}$, as shown in Fig.~\ref{fig:m_corot} and Fig.~\ref{fig:a_corot}. However, as pointed out earlier, the impactor flux $F_{imp}$ we use could be orders of magnitude overestimated, and there are also uncertainties on the yields $Y$. This means the amount of dust in the arc can be at least an order of magnitude smaller.
The predicted optical depth is much smaller than the total optical depth of the E ring at this region, but the E ring is dominated by 1$\mum$ particles while larger particles dominate the optical depth of the arcs. Thus, this could explain the observability of the arcs in low phase images \citep{Hedman:2009kt}.

On the other hand, we ignored a potential source of dust. Based on the absorption of electrons observed by Cassini Low Energy Magnetospheric Measurement System (LEMMS), \citet{Roussos:2008bi} proposed an arc composed of particles larger than a millimeter near Methone, and the optical depth is below the detection limit. These `large' particles could also be the source of smaller ones via impact-ejecta process. Assuming the optical depth of these larger particles is $10^{-7}$, and assume the 24$^\circ$ Methone arc center at radial distance of Methone ($194170\km$) and the width is 1000$\km$, then the total cross section of large particles is about 8.1$\km^2$. This is slightly larger than the cross section of Methone (${\sim}6.6\km^2$). Since the dust production rate is proportional to the cross section (Eq.~\eqref{eq:mplus}) as long as all ejecta can escape their source, this implies the source rate can be doubled.
Additionally, particles larger than 30$\mum$ could also contribute to the optical depths due to their longer lifetimes which has not been considered in the calculations.

\subsection{Summary}
We use numerical simulations to estimate geometric optical depth of Methone's and Anthe's arcs. To do this, we model the source, the dynamical evolution, and the sinks of particles. We use the impact-ejecta model \citep{Krivov:2003ku, Spahn:2006dh} to estimate the source rate of particles, though the parameters, especially $F_{imp}$ and $Y$ should need further updates. For the dynamical evolution, we consider the perturbing forces: oblateness of Saturn, solar radiation pressure, Lorentz force, and plasma drag. Mimas and Enceladus are both considered: Mimas play an important role in confining the arc material, while Enceladus is the next massive moon outside these two arcs.
The dominating mechanism to remove particles from the arc is plasma drag which pushes them outwards. About 10-20\% of all particles end on Enceladus. If only consider the larger particles ($s>10\mum$), recollision with the source moons is as important as migrating outward.
The erosion of particles by plasma sputtering is also taken into account, which not only limits the maximum lifetime of particles, but also implies the dynamics change with time because of the decreasing grain radii.

\section*{Acknowledgments}
    We appreciate Matt Hedman for the fruitful discussions.
    The work is supported by DFG SP 384/21-1.


\end{document}